# Frequency driven inversion of tunnel magnetoimpedance in magnetic tunnel junctions

Subir Parui[1,*], Mário Ribeiro[1], Ainhoa Atxabal[1], Amilcar Bedoya-Pinto[1,2], Xiangnan Sun[1,3], Roger Llopis[1], Fèlix Casanova[1,4], Luis E. Hueso[1,4,*]

[1] *CIC nanoGUNE, 20018 Donostia-San Sebastian, Spain*

[2] *Max Planck Institute of Microstructure Physics, D-06120 Halle, Germany*

[3] *National Center for Nanoscience and Technology 100190 Beijing, P. R. China*

[4] *IKERBASQUE, Basque Foundation for Science, 48011 Bilbao, Spain*

Email: s.parui@nanogune.eu; l.hueso@nanogune.eu

KEYWORDS: Tunnel barrier, spintronics, magnetic tunnel junction

## Abstract

Magnetic tunnel junctions (MTJs) are basic building blocks for devices such as magnetic random access memories (MRAMs). The relevance for modern computation of non-volatile high-frequency memories makes ac-transport measurements of MTJs crucial for exploring this regime. Here we demonstrate a frequency-mediated effect in which the tunnel magnetoimpedance reverses its sign in a classical $Co/Al_2O_3/NiFe$ MTJ, whereas we only observe a gradual decrease of tunnel magnetophase. Such effects are explained by the capacitive coupling of a parallel resistor and capacitor in the equivalent circuit model of the MTJ. Furthermore, we report a positive tunnel magnetocapacitance effect, suggesting the presence of a spin-capacitance at the two ferromagnet/tunnel-barrier interfaces. Our results are important for understanding spin transport phenomena at the high frequency regime, in which the spin-polarized charge accumulation at the two interfaces plays a crucial role.

# I. Introduction

Magnetic random access memory (MRAM) devices based on magnetic tunnel junctions (MTJ) are very attractive technologically, mostly due to their low energy consumption and fast data processing [1-5]. The two-terminal geometry of the MTJs is based on a thin dielectric layer (I) acting as the tunnel barrier, sandwiched between two ferromagnetic electrodes (FM), in a FM/I/FM structure [1-5]. MTJs typically show low (/high) resistance states when the magnetic orientation of the two ferromagnetic electrodes is parallel (/antiparallel), providing a positive tunnel magnetoresistance (TMR) [1-7]. Beyond this conventional positive TMR effect there are several non-trivial effects that give rise to negative TMR, where the low (/high) resistance states now correspond to the antiparallel (/parallel) orientation of the electrodes magnetization, namely spin-polarized resonance [8], quantum size effects [9], modulation of the density-of-states (DOS) of the ferromagnetic electrodes [10], inversion of tunneling spin polarization at the FM/I interface [11], and the formation of a tunneling standing wave [12]. These mechanisms are usually addressed by either dc (direct current) or ac (alternating current)-transport measurements at low frequencies [12, 13]. For ac measurements, the impedance (Z), i.e., the total dynamic resistance of the circuit converges to the resistance measured only with dc-voltages when the phase shift ($\theta$) is negligible. Phenomena such as spin polarized resonant tunneling, in which the TMR changes from positive to negative, is crucial for the development of resonant-tunneling spin transistors and quantum information devices [8]. However, so far this effect has been shown to occur only when the thickness of a metallic insertion layer (Cu) between the FM and I layers is varied within a certain range [8]. Accordingly, it would be important to have alternative ways for controlling the magnetoimpedance (or magnetoresistance) externally that does not depend on the structural characteristics of the device.

Here, we demonstrate a controlled positive-to-negative inversion of the tunnel magnetoimpedance with increasing frequencies in a classical Co/Al$_2$O$_3$/NiFe MTJ by using impedance-spectroscopy measurements. The magnitude of the inversion depends on the individual resistance change between parallel and antiparallel magnetic states and the change in the spin-capacitance [14] of the FM/I interfacial layer. Simultaneously, the impedance-spectroscopy measurements not only help us to determine the presence of spin capacitance but also to characterize the speed of the device operation, and to estimate the tunnel magnetocapacitance effect [15-19]. We report a positive tunnel magnetocapacitance for the first time using a simple parallel equivalent circuit of a resistor and a capacitor of the MTJ, and by measuring the impedance and phase at set frequencies. In the regime of frequencies where the equivalent parallel circuit model is reliable, the tunnel magnetoresistance measured by ac-method comes in agreement with the standard dc measurements. Our results open the possibility to exploit the impedance-spectroscopy measurements for the optimization of functional spintronic devices.

## II. Experimental details

A vertical cross-bar geometry of Co (16 nm)/$Al_2O_3$ (x nm)/$Ni_{80}Fe_{20}$ (16 nm) (from bottom to top, see Figure 1) was deposited *in situ* through different shadow masks. The junction stack is fabricated on a Si/$SiO_2$ (150 nm) substrate. Initially, several 16-nm-thick Co bars are deposited by e-beam evaporation in an ultrahigh vacuum (UHV) chamber at the rate of 1Å/s. Afterwards, the $Al_2O_3$ tunnel barrier is obtained evaporating Al (2.5 or 3 nm) all over the device, followed by plasma oxidation. Two steps of plasma oxidation at the oxygen pressure of $10^{-1}$ mbar are employed in order to obtain a robust $Al_2O_3$ tunnel barrier [20]. Finally, a 16-nm-thick NiFe top electrode was deposited at the rate of 1Å/s through a different shadow mask. The active area for tunneling is considered to be confined to the overlapping area between the Co bar and the NiFe bar (here we report results from two different device areas of 0.5 and 0.2 $mm^2$). The $Al_2O_3$ dielectric sandwiched between the two ferromagnetic electrodes also provides a two terminal capacitor. Once fabricated, the device is transferred into a variable-temperature probe station (Lakeshore) for electrical measurements with a Keithley-4200 CVU semiconductor analyzer, which is capable to carry out both the impedance-spectroscopy measurement in the frequency range of $10^3$-$10^7$ Hz and the pure dc measurements. Using a two-probe configuration we measure both types of transport characteristics using the same two probes to avoid any circuit complication in the high frequency operation [19]. In Figure 1, the dotted line (red) shows the standard dc measurement schematics and the solid line (black) represents the schematics of the impedance-spectroscopy (ac + dc) measurement.

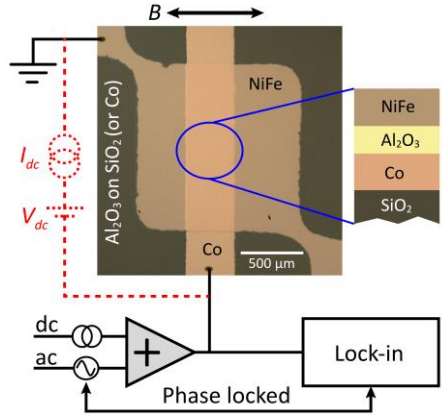

FIG. 1 Optical microscopy image of the device and the schematic diagram of the MTJ of Co/$Al_2O_3$/NiFe stacked on a Si/$SiO_2$ substrate. The dotted red-line represents the standard two-probe dc measurement schematics and the solid black-line represents the schematics of the impedance-spectroscopy measurement. Sweeping direction of the applied magnetic field *B* is also shown.

# III. Results and discussions

**A. DC-transport characteristics**

We first focus on the dc-transport measurements of the MTJs. In case of a thin tunnel barrier, the s- and d-orbital electrons that tunnel from one ferromagnet to the other conserve the spin information via the single-step tunnel process [21]. The transport can be described as having two independent spin channels of two different resistances. In general, a low resistance state is observed for parallel alignment ($R_P$) of the two ferromagnets, and a high resistance state ($R_{AP}$) is observed when they are anti-parallel. The associated tunneling magnetoresistance ($TMR_{DC}$) ratio is defined as

$$TMR_{DC} = \frac{(R_{AP}-R_P)}{R_P} \times 100. \qquad (1)$$

The electric resistance of the parallel and anti-parallel magnetic states of the device is measured by two-probe method with the applied dc voltage up to ±600 mV.

Figure 2a shows the $TMR_{DC}$ data for a representative Co/Al$_2$O$_3$ (oxidized from 3 nm Al)/Ni$_{80}$Fe$_{20}$ MTJ. When a large enough magnetic field is applied in-plane, the magnetizations of the NiFe and Co electrodes are aligned in parallel. Sweeping the magnetic field, causes the magnetization of NiFe to switch at a much lower field than that of Co, and an antiparallel state is achieved. The magnetization of Co and NiFe can thus be aligned either parallel (P) or antiparallel (AP), and two distinct resistance states are observed as can be seen in the measurement. The $TMR_{DC}$ in the MTJ reaches 20% at the lowest temperatures, at +10 mV of dc bias (Figure 2a). This value is comparable to previous literature [1, 7, 22], confirming the good quality of our MTJs.

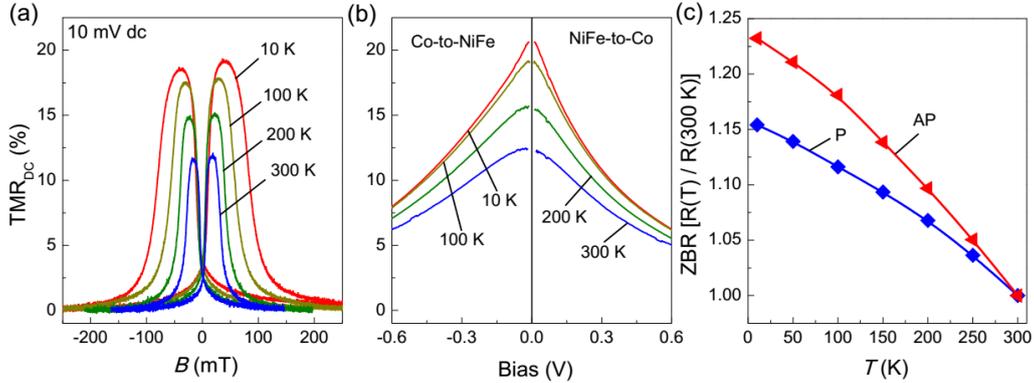

FIG. 2 (a) Spin-valve effect at several temperatures for the Co/Al$_2$O$_3$/NiFe MTJ at a dc bias of +10 mV. The bias is applied at the Co electrode with respect to the grounded NiFe electrode in two-probe geometry. (b) $TMR_{DC}$ as a function of the applied bias at several temperatures. (c) Zero bias resistance (ZBR) vs temperature for P and AP magnetization orientations of the two ferromagnets of the MTJ. The ZBR exhibits modest temperature dependence, confirming direct tunneling.

Furthermore, we have obtained the measurements of $TMR_{DC}$ by sweeping the bias voltage and measuring individually the resistance corresponding to either in the P-state or in the AP-state. Figure 2(b) shows that the magnitude of the TMR$_{DC}$ decreases with increasing bias in a monotonic, nonlinear fashion at all temperatures from 300 K to 10 K. This is a standard behavior for MTJs that relies on the basic physics of the spin-polarized tunneling mechanism. Increasing the bias voltage increases the spin relaxation rate, which decreases the TMR$_{DC}$ as the electrons tunnel into empty states of the receiving electrode with an excess energy, generating phonons and magnons [22, 23]. The asymmetric bias dependence of TMR$_{DC}$ reflects the different DOS of the Co and NiFe metals used. At the same time, decreasing the temperature reduces the phonon and magnon scattering, leading to an increase in the TMR$_{DC}$. Figure 2(c) presents the temperature dependence of the zero bias resistance (ZBR) of the tunnel barrier, defined as R(T)/R(300 K). It exhibits modest temperature dependence, characteristics of the direct tunneling process between the two ferromagnetic materials.

**B. AC-transport characteristics**

In order to investigate the tunnel magnetoimpedance effect, we focus on the impedance-spectroscopy measurement combining an ac bias with an applied dc bias. From the ac-transport measurement, we obtain the impedance and the phase (Z, θ) as the fundamental parameters. From Z and θ, we can obtain $Re(Z) = |Z|\cos\theta$ and $Im(Z) = |Z|\sin\theta$; where $Z = Re(Z) + iIm(Z)$ and $|Z| = \sqrt{Re(Z)^2 + Im(Z)^2}$. We must note that both Z and Re(Z) converge to the dc value of the resistance in case of θ→0 at the low frequency regime, so that |Z| ≈ Re(Z) ≈ R$_{DC}$.

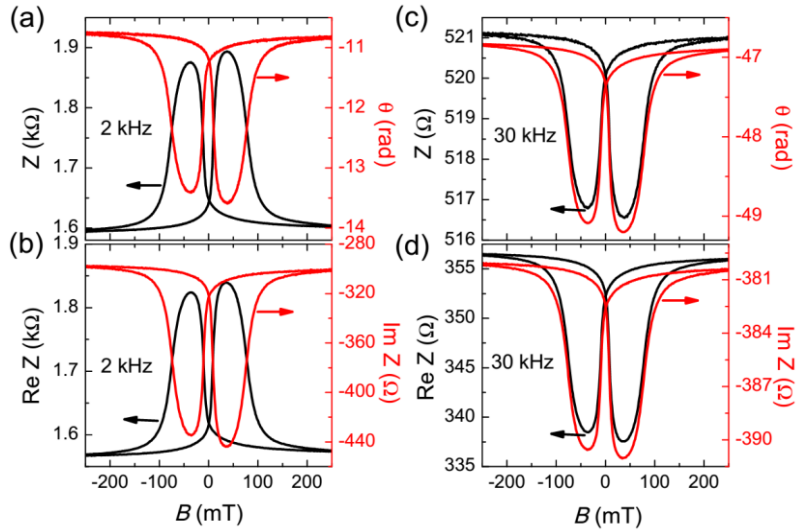

FIG. 3 (a), (b) Magnetic field dependent (Z, θ), and [Re(Z), Im(Z)], respectively, at 2 kHz. Measurements are performed individually at +10 mV of dc bias, 30 mV of RMS ac bias and at 10 K. (c), (d) Similar measurements as in (a) and (b) are now performed at 30 kHz. At this frequency, the switching of Z and Re(Z) are inverted.

Figure 3 represents (Z, θ) and [Re(Z), Im(Z)] as a function of the magnetic field for two different frequencies, 2 kHz and 30 kHz, respectively, for the same MTJ as discussed in the dc-transport measurement. The black (left axis) and the red (right axis) arrows are used to indicate Z, Re(Z) and θ, Im(Z) respectively. Comparing these two frequencies, we see that the switching of both Z and Re(Z) (black curves) becomes inverted, whereas θ and X (red curves) remain consistent with their usual switching behavior. It is worth noting that the overall Z and Re(Z) value decreases at 30 kHz compared to the frequency at 2 kHz. However, an increase of θ and Im(Z) suggests the presence of the capacitive coupling in the device. Nevertheless, the switching behaviors of Z and Re(Z) with respect to the magnetic field clearly confirm the frequency-driven inversion of tunnel magnetoimpedance in the MTJ.

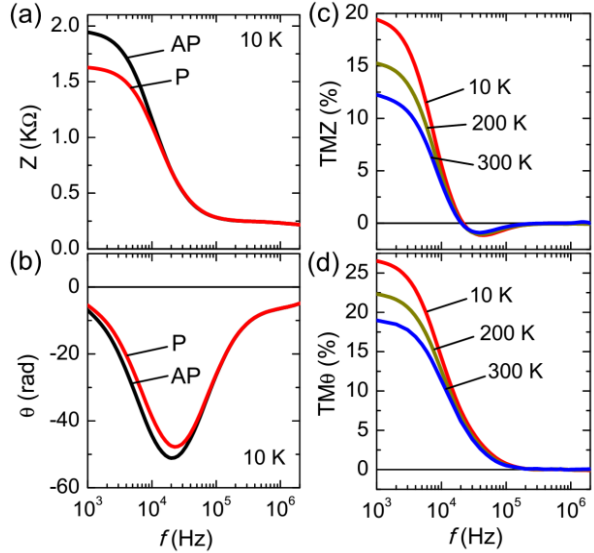

FIG. 4 (a) Impedance and (b) phase as a function of frequency at 10 K for the P and AP orientations of the magnetization of the two FM electrodes. All the two-probe impedance-spectroscopy measurements are carried out at dc bias of +10 mV and at RMS ac bias of 30 mV. (c) Tunneling magnetoimpedance (TMZ) and (d) tunneling magnetophase (TMθ) as a function of frequency at different temperatures.

In order to identify the different frequency regimes for positive and negative tunnel magnetoimpedance, we performed frequency scans of Z and θ for parallel and antiparallel magnetic orientations, respectively. Figure 4 shows the frequency-dependence of Z, θ, TMZ, and TMθ respectively. We calculate TMZ and TMθ by using the same formula as in equation (1) but by changing the respective parameters (Z and θ). Figure 4(c) shows that the TMZ converges to the $TMR_{DC}$ values in the low frequency regime of ~1 kHz. Due to the capacitive coupling that originates from the charge accumulation at the two interfaces between the ferromagnets and the $Al_2O_3$ dielectric, both the TMZ and TMθ values decrease with increasing frequency. The damping of the frequency response depends on the frequency independent resistance and the capacitance of the MTJ, which can also be further controlled by engineering the MTJs, as will be discussed below. Nevertheless, it is worth noting that the TMZ values [see Figure 4(c)] flip their sign in the frequency range of ~$2\times10^4$ to ~$2\times10^5$ Hz before reaching zero. We propose that the inversion of TMZ is related to the quicker damping of Z in the AP state relative to the P state as a function of the frequency. This can be understood on the basis of different resistances and capacitances in the AP and P state resulting in different time constants, which eventually control the damping of Z and of its inversion.

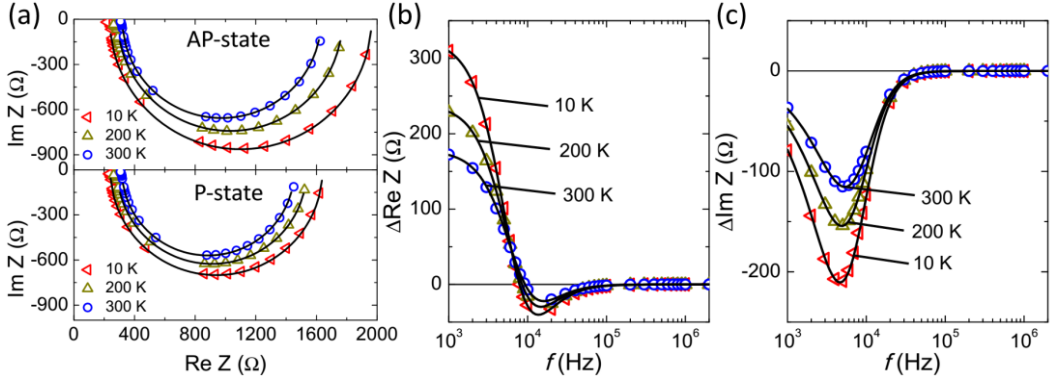

FIG. 5 (a) Cole-Cole plots of the MTJ for the frequency range of 1 kHz-2 MHz in P and AP-state. The semicircular fits (solid black line) match very well with the experimental data. (b) $\Delta Re(Z)$ is shown as a function of the frequency. The solid black line is the fit by using equation (2) and the respective fitting parameters are listed in the Table I. (c) $\Delta Im(Z)$ is shown as a function of the frequency. The solid black line is the fit by using equation (3) with the same fitting parameters as before.

It is possible to interpret the impedance-spectroscopy measurements via an equivalent circuit model. In MTJs, the resistance and capacitance are coupled in a parallel equivalent circuit [15-19]. Figure 5(a) represents the Cole-Cole diagrams. The plots between Re(Z) and Im(Z) show semicircular dependences, which validate the parallel circuit model [18, 19]. In order to simplify the circuit analysis and to determine the respective frequency (f), independent resistance ($R_f$) and capacitance ($C_f$), the magnetic field independent part during ac-measurement can be eliminated by taking the difference of the impedances in the P and AP states [18]. This difference can be written as $\Delta Z = \Delta Re(Z) + i\Delta Im(Z)$, where

$$\Delta Re(Z) = \left(\frac{R_f(AP)}{1+4\pi^2 f^2 R_f^2(AP) C_f^2(AP)} - \frac{R_f(P)}{1+4\pi^2 f^2 R_f^2(P) C_f^2(P)}\right), \quad (2)$$

$$\Delta Im(Z) = -2\pi f \left(\frac{C_f(AP) R_f^2(AP)}{1+4\pi^2 f^2 R_f^2(AP) C_f^2(AP)} - \frac{C_f(P) R_f^2(P)}{1+4\pi^2 f^2 R_f^2(P) C_f^2(P)}\right). \quad (3)$$

Figures 5(b) and 5(c) show the frequency dependence of the $\Delta Re(Z)$ and the $\Delta Im(Z)$ of the impedance. The frequency independent fitting parameters $R_f(AP)$, $R_f(P)$, $C_f(AP)$, and $C_f(P)$ are extracted by using both equations (2) and (3), and are listed in the Table I. From these parameters we observe that the positive tunnel magnetoresistance (~23% at 10 K) is consistent with the dc measurements, whereas almost no tunnel magnetocapacitance effect is extracted. This result coincides with a previous report [17], although it is fair to admit the conflicting studies in this regard [15, 16, 18, 19].

The temperature independent capacitance values extracted from these fittings amount to 12.7 nF. By considering the parallel plate capacitor device structure of our MTJs, we calculate the geometrical-capacitance ($C_g = \varepsilon \frac{A}{d}$) to be 10.2 nF, where $A$ is the area of the junction (0.5 mm$^2$), $d$ is the thickness (3-nm-thick Al yields an Al$_2$O$_3$ tunnel barrier of 3.9 nm as $d_{Al_2O_3} \approx 1.3 d_{Al}$ [24]), and $\varepsilon$ is the dielectric constant of Al$_2$O$_3$ ($9\varepsilon_0$, $\varepsilon_0 = 8.85 \times 10^{-12}$ F/m). The capacitance extracted from the fit is slightly larger than the geometrical capacitance, which suggests the presence of an additional leaky capacitance in parallel to the geometrical capacitance [18], making the model unreliable for the detection of interface spin capacitance. Below we discuss a straightforward approach to detect the influence of the two interface spin-capacitances which are in principle in series with the geometrical capacitance [16, 18], and give rise to the positive tunnel magnetocapacitance.

TABLE I. Extracted fitting parameters from Figures 5(b) and 5(c), and the corresponding time constants.

| T | $R_f(AP)$ | $R_f(P)$ | $C_f$ (AP) | $C_f$ (P) | $R_f C_f$(AP) | $R_f C_f$(P) |
|---|---|---|---|---|---|---|
| 10 K | 1725±10 Ω | 1400±9 Ω | 12.7±0.5 nF | 12.7±0.5 nF | 21.9±0.9 μs | 17.8±0.7 μs |
| 200 K | 1604±8 Ω | 1365±8 Ω | 12.7±0.4 nF | 12.7±0.4 nF | 20.3±0.6 μs | 17.3±0.5 μs |
| 300 K | 1448±9 Ω | 1270±8 Ω | 12.7±0.4 nF | 12.7±0.4 nF | 18.4±0.6 μs | 16.1±0.5 μs |

In a parallel resistor-capacitor circuit model, the time constant is determined as the product of the corresponding resistance and capacitance. Such time constant of MTJs plays an important role for high-speed applications. From our large area tunnel junction, we determine the time constants to be of the order of μs. However, the time constant of the circuit changes in two different magnetic configurations (P and AP) as well as in different temperatures due to the change in resistance in the circuit. A cutoff frequency $f_c = \frac{1}{2\pi R_f C_f}$ can thus be obtained. For example, the cut-off frequencies at 10 K are 7.3 kHz (AP-state) and 8.9 kHz (P-state) respectively. Interestingly, these cut-off frequencies are quite close to the frequency regime above which we observe the inversion of TMZ as can be seen in figures 4(c).

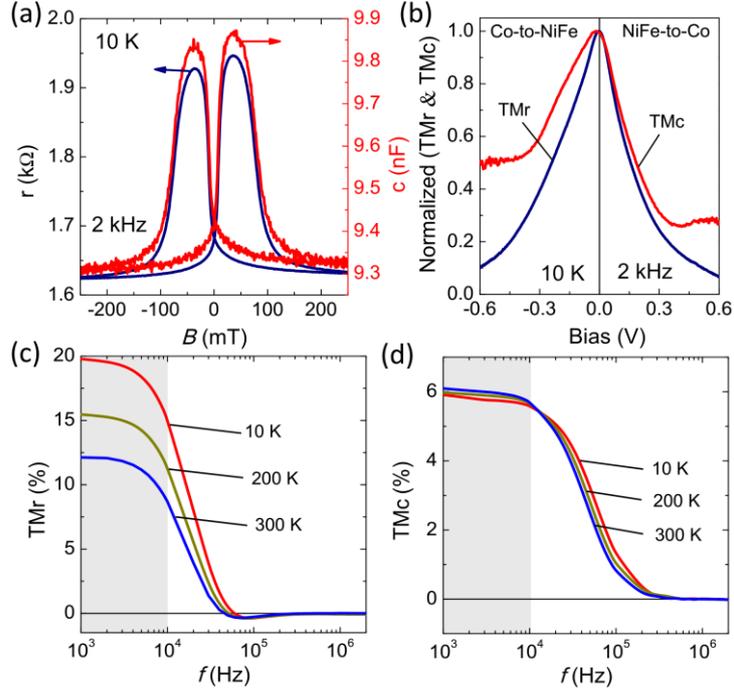

FIG. 6 (a) The magnetic field dependence of resistance (r) and capacitance (c) of the MTJ by considering the parallel equivalent circuit. The measurements are done at dc bias of +10 mV and at RMS ac bias of 30 mV. (b) Normalized bias dependence of both TMr and TMc of the device. Frequency dependence of the TMr (c) and TMc (d) are shown at different temperatures.

Further, we perform a more straightforward analysis to determine the resistance and the capacitance [15, 16]. Considering the MTJ as a parallel network of r and c, we calculate their values directly at each set frequency from the (Z, θ) measurements. The direct dependence can be expressed as follows, $r = Z\sqrt{1 + (\tan\theta)^2}$ and $c = (Z\omega)^{-1}[1 + (\tan\theta)^{-2}]^{-1/2}$, where $\omega = 2\pi f$ is the frequency of the ac signal. Figure 6(a) shows the typical c and r dependence with varying magnetic fields. Our measurement shows a positive TMc value, smaller than the positive TMr values. The positive value of TMc indicates that there is higher capacitance in the AP-state than the P-state. To the best of our knowledge, this is the first time where a positive TMc is observed in a MTJ [15, 16, 18, 19]. We suggest that one of the possible mechanisms behind the positive TMc is the different spin-dependent screening length of the spin polarized charge accumulation at the two different FM/I interfaces (Co/Al$_2$O$_3$ and NiFe/Al$_2$O$_3$). Figure 6(b) reveals the dc bias dependence of the normalized TMr and TMc during the impedance-spectroscopy measurements. We observe a decrease with increasing bias for both magnitudes, which is consistent with standard dc measurements. We also observe that the TMc is relatively less sensitive to the applied dc bias than the TMr, indicating a low leakage current in the MTJ [18]. Figures 6(c) and 6(d) show the frequency dependence of the TMr and TMc at different temperatures. In an ideal case, TMr and TMc are expected to be frequency independent. However, TMr and TMc decrease very sharply after a certain frequency, and we also notice a weak inversion in TMr. Nevertheless, in the low frequency regime we observe an almost temperature independent TMc, whereas the TMr is very consistent with the standard dc-measurements at different temperatures.

An in-plane magnetic field dependence of the capacitance of a parallel plate capacitor can be induced via a combination of the Lorentz force and quantum confinement effects [25, 26] or by spin-dependent Zeeman splitting [27, 28]. Such phenomena are named as *magnetocapacitance* effect in a system without any ferromagnetic electrode. However, the origin of tunnel magnetocapacitance in MTJs can only be attributed to the presence of spin-capacitance at the two FM/I interface that appears due to the capacitive accumulation of spin-polarized carriers at the interface [14].

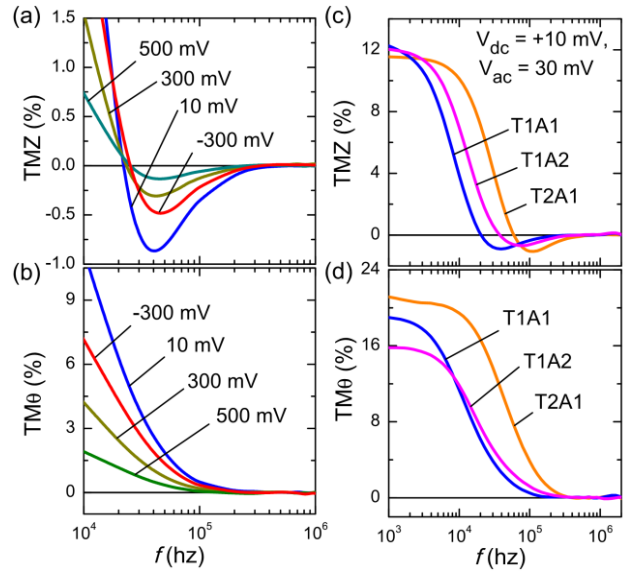

FIG. 7 (a), (b) TMZ and TMθ at various dc bias voltages as a function of frequency at RMS ac bias of 30 mV. (c), (d) TMZ and TMθ as a function of frequency for two different thickness of $Al_2O_3$ (T1A1: 3 nm and T2A1: 2.5 nm but with same junction areas) as well as for two different device areas (T1A1: 0.5 mm$^2$ and T1A2: 0.2 mm$^2$) with the same thickness of the $Al_2O_3$. All measurements are performed at room temperature.

In order to have further control over the inversion phenomena in the tunnel magnetoimpedance, it is thus also important to have a different frequency response of the MTJs. Since the tunnel resistance is bias dependent, it is expected a change in the frequency response and also in the amplitude of inversion of the tunnel magnetoimpedance. Figures 7(a) and 7(b) show the frequency response of TMZ and TMθ at several dc-biases, whereas the ac-voltage is kept constant. Plots of the TMZ show a sharply damped oscillation and an increase in the frequency response of damping with increasing dc-bias. However, none of the TMθ plots show inversion. Furthermore, a modification in the thickness of the tunnel barrier and/or the area of the overlapping electrodes in the MTJ can be used to tune the damping frequency. Figures 7(c) and 7(d) show the frequency response of TMZ and TMθ of two different tunnel barrier thicknesses, which are oxidized from 3 nm (T1A1) and 2.5 nm (T2A1) thick Al, but with same overlapping area between Co and NiFe electrodes. In case of the thinner tunnel barrier, the resistance of the junction decreases exponentially with decreasing thickness, whereas the capacitance increases, since it is inversely proportional to the thickness between the two electrodes. Thus, for the tunnel barrier from 2.5-nm-thick Al, we observe the cut off frequency of the inversion increases compared to the junction with the tunnel barrier from 3-nm-thick Al and our observation of frequency shift is in agreement with the previous literature [16]. The figures also include the frequency response of the MTJs with different areas of the two overlapping electrodes (T1A1: 0.5 mm$^2$ and T1A2: 0.2 mm$^2$) for the same thickness of the tunnel barrier (oxidized from 3-nm-thick Al). Both the tunnel resistance and the capacitance change with the effective device area and we observe higher frequency shift of the inversion with decreasing device area. Consequently, we confirm that the inversion in tunnel magnetoimpedance is very robust and easily tunable by applied dc bias and/or by engineering the junction geometrical parameters.

## IV. Conclusions

In conclusion, we demonstrate an alternative mechanism to achieve inversion in the tunnel magnetoimpedance of MTJs, based on the frequency of the applied ac-signal. Ac-transport measurements of the impedance and corresponding phase allow the extraction of other fundamental parameters, such as resistance and capacitance. From the straightforward determination of the equivalent resistance and capacitance, we determine not only a positive tunnel magnetoresistance but also a positive tunnel magnetocapacitance effect. Moreover, in order to establish the robustness of the inversion phenomena, we report the dependence of the frequency response of the magnetoimpedance with the dc bias-voltage, thickness of the tunnel barrier and area of the two overlapping electrodes of the MTJs. Our results are encouraging to understand the spin-polarized charge accumulation as well as spin transport phenomena in spintronic devices at high frequencies involving any thin dielectric tunnel barrier, including ferromagnetic insulators. In particular, the inversion phenomena in MTJs will help extending the current use of Al$_2$O$_3$ tunnel barriers in hot electron transistors [29, 30] as well as encourage the development of highly functional spin transport devices.

## Acknowledgements

We are grateful to Prof. Hanan Dery for useful discussions. We also acknowledge financial support from the European Union's 7th Framework Programme under the European Research Council (Grant 257654-SPINTROS), under People Programme (Marie Curie Actions) REA grant agreement 607904-13, and under the NMP project

**References**

[1] J.S. Moodera, L.R. Kinder, T.M. Wong, and R. Meservey, Phys. Rev. Lett. **74**, 3273 (1995).

[2] S.S.P. Parkin et al., Proc. IEEE **91**, 661 (2003).

[3] S. S. P. Parkin et al., Nat. Mater. **3**, 862 (2004).

[4] S. Yuasa et al., Nat. Mater. **3**, 868 (2004).

[5] S.A. Wolf et al., Science **294**, 1488 (2001).

[6] E. Cobas, A.L. Friedman, O.M.J. van't Erve, J.T. Robinson, and B.T. Jonker, Nano Lett. **12**, 3000 (2012).

[7] A. Bedoya-Pinto, M. Donolato, M. Gobbi, L. E. Hueso, and P. Vavassori, Appl. Phys. Lett. **104**, 062412 (2014).

[8] S. Yuasa et al., Science **297**, 234, (2002).

[9] J. S. Moodera et al., Phys. Rev. Lett. **83**, 3029, (1999).

[10] P. LeClair, B. Hoex, H. Wieldraajjer, J. T. Kohlhepp, H. J. M. Swagten, and W. J. M. de Jonge, Phys. Rev. B **64**, 100406(R) (2001).

[11] I. J. Vera Marún, F. M. Postma, J. C. Lodder, and R. Jansen, Phys. Rev. B **76**, 064426 (2007).

[12] C. W. Miller, Z. -P. Li, I. K. Schuller, R. W. Dave, J. M. Slaughter, and J. Åkerman, Phys. Rev. Lett. **99**, 047206 (2007).

[13] L. Gao, X. Jiang, S. H. Yang, J. D. Burton, E. Y. Tsymbal, and S. S. P. Parkin, Phys. Rev. Lett. **99**, 226602 (2007).

[14] J. M. Rondinelli, M. Stengel, and N. A. Spaldin, Nature Nanotech. **3**, 46 (2008).

[15] H. Kaiju, S. Fujita, T. Morozumi, and K. Shiiki, J. Appl. Phys. **91**, 7430 (2002).

[16] P. Padhan, P. LeClair, A. Gupta, K. Tsunekawa, and D. Djayaprawira, Appl. Phys. Lett. **90**, 142105 (2007).

[17] S. Ingvarsson, M. Arikan, M. Carter, W. F. Shen, and Gang Xiao, Appl. Phys. Lett. **96**, 232506 (2010).

[18] A. M. Sahadevan, K. Gopinadhan, C. S. Bhatia, and H. Yang, Appl. Phys. Lett. **101**, 162404 (2012).

[19] Y.-M. Chang, K.-S. Li, H. Huang, M.-J. Tung, S.-Y. Tong, and M.-T. Lin, J. Appl. Phys. **107**, 093904 (2010).

[20] P. LeClair, J. T. Kohlhepp, A. A. Smits, H. J. M. Swagten, B. Koopmans, and W. J. M. de Jonge, J. Appl. Phys. **85,** 7803 (1999); K. S. Yoon, J. H. Park, J. H. Choi, J. Y. Yang, C. H. Lee, C. O. Kim, J. P. Hong, and T. W. Kang, Appl. Phys. Lett. **79,** 1160 (2001).

[21] M. Jullière, Phys. Lett. **54A**, 225 (1975).

[22] B. G. Park, T. Banerjee, J. C. Lodder, and R. Jansen, Phys. Rev. Lett. **99**, 217206 (2007).

[23] S. Zhang, P. M. Levy, A. C. Marley, and S. S. P. Parkin, Phys. Rev. Lett. **79**, 3744 (1997).

[24] G. Landry, Y. Dong, J. Du, X. Xiang, and John Q. Xiao, Appl. Phys. Lett. **78**, 501 (2001).

[25] J. Hampton, J. Eisenstein, L. Pfeiffer, and K. West, Solid State Commun. **94**, 559 (1995).

[26] T. Jungwirth and L. Smrčka, Phys. Rev. B **51**, 10181 (1995).

[27] S. Zhang, Phys. Rev. Lett. **83**, 640 (1999).

[28] K. T. McCarthy, A. F. Hebard, and S. B. Arnason, Phys. Rev. Lett. **90**, 117201 (2003).

[29] M. Gobbi, L. Pietrobon, A. Atxabal, A. Bedoya Pinto, X. Sun, F. Golmar, R. Llopis, F. Casanova, and L. E. Hueso, Nat. Commun. **5,** 4161 (2014).

[30] S. Parui, A. Atxabal, M. Ribeiro, A. Bedoya-Pinto, X. Sun, R. Llopis, F. Casanova and L. E. Hueso, Appl. Phys. Lett. **107,** 183502 (2015).